\documentclass[a4paper]{jpconf}
\usepackage{graphicx}
\usepackage{braket}
\usepackage{subfigure}
\usepackage{url}
\usepackage{cite}
\usepackage{hyperref} 
\begin{document}
\title{A Full Quantum Generative Adversarial Network Model for High Energy Physics Simulations}

\author{Florian Rehm$^{1, 2}$, Sofia Vallecorsa$^{1}$, Michele Grossi$^1$, Kerstin Borras$^{2, 3}$, Dirk Krücker$^3$
}
\address{$^1$ CERN, 1211 Geneva 23, Switzerland}
\address{$^2$ RWTH Aachen University, Templergraben 55, 52062 Aachen, Germany}
\address{$^3$ Deutsches Elektronen-Synchrotron DESY, Notkestrasse 85, 22607 Hamburg, Germany}
\ead{florian.matthias.rehm@cern.ch\\ 
}

\begin{abstract}
The prospect of quantum computing with a potential exponential speed-up compared to classical computing identifies it as a promising method in the search for alternative future High Energy Physics (HEP) simulation approaches. HEP simulations, such as employed at the Large Hadron Collider at CERN, are extraordinarily complex and require an immense amount of computing resources in hardware and time. For some HEP simulations, classical machine learning models have already been successfully developed and tested, resulting in several orders of magnitude speed-up. In this research, we proceed to the next step and explore whether quantum computing can provide sufficient accuracy, and further improvements, suggesting it as an exciting direction of future investigations.

With a small prototype model, we demonstrate a full quantum Generative Adversarial Network (GAN) model for generating downsized eight-pixel calorimeter shower images. The advantage over previous quantum models is that the model generates real individual images containing pixel energy values instead of simple probability distributions averaged over a test sample. 

To complete the picture, the results of the full quantum GAN model are compared to hybrid quantum-classical models using a classical discriminator neural network.
\end{abstract}

\section{Introduction}
\label{ch:fullQGAN}
The full quantum Generative Adversarial Network (qGAN) model consists of a generator and a discriminator, both of which are variational quantum circuits \cite{variational_quantum_circuits} with trainable parameters. While hybrid quantum-classical GAN models have been well studied, successfully operating full qGAN models are still rare in the literature due to the complexity of the implementation and difficulties in training towards convergence. 
For example, Ref. \cite{qGAN_example} proposed a full qGAN model that can generate quantum states with a high fidelity. Another work, Ref. \cite{qGAN_example2} introduced a full qGAN that can be trained with a limited number of samples.

The use case to which we apply the full qGAN model in this study is the calorimeter detector simulation, employed in HEP researches to measure the energies of particles \cite{calorimeter_cern}. The preprocessed training and test data sets contain both approximately 1\,000 simplified eight pixel images and are available at Ref. \cite{quantum_data_set}. The data sets were generated for a proposed future CLIC calorimeter \cite{CLIC_dataset} by the Geant4 tool \cite{Geant4} for primary particle energies between $E \in [225, 275]\,$GeV. The model utilizes angle encoding \cite{state_preparation2}, which has a linear scaling from the number of image pixels to the number of qubits. Here, eight qubits are required in the quantum circuits.

This paper comprises: an introduction to the model and the quantum circuit architecture, the encoding and decoding, and the training parameters. Then, the full qGAN model is evaluated in training and inference. Finally, the full qGAN is compared to a hybrid model to compare the expressive power of quantum circuits to classical neural networks.

\section{Description of the Full Quantum GAN Model}
\begin{figure}[t!]  
    \center
    \includegraphics[height=0.32\textwidth]{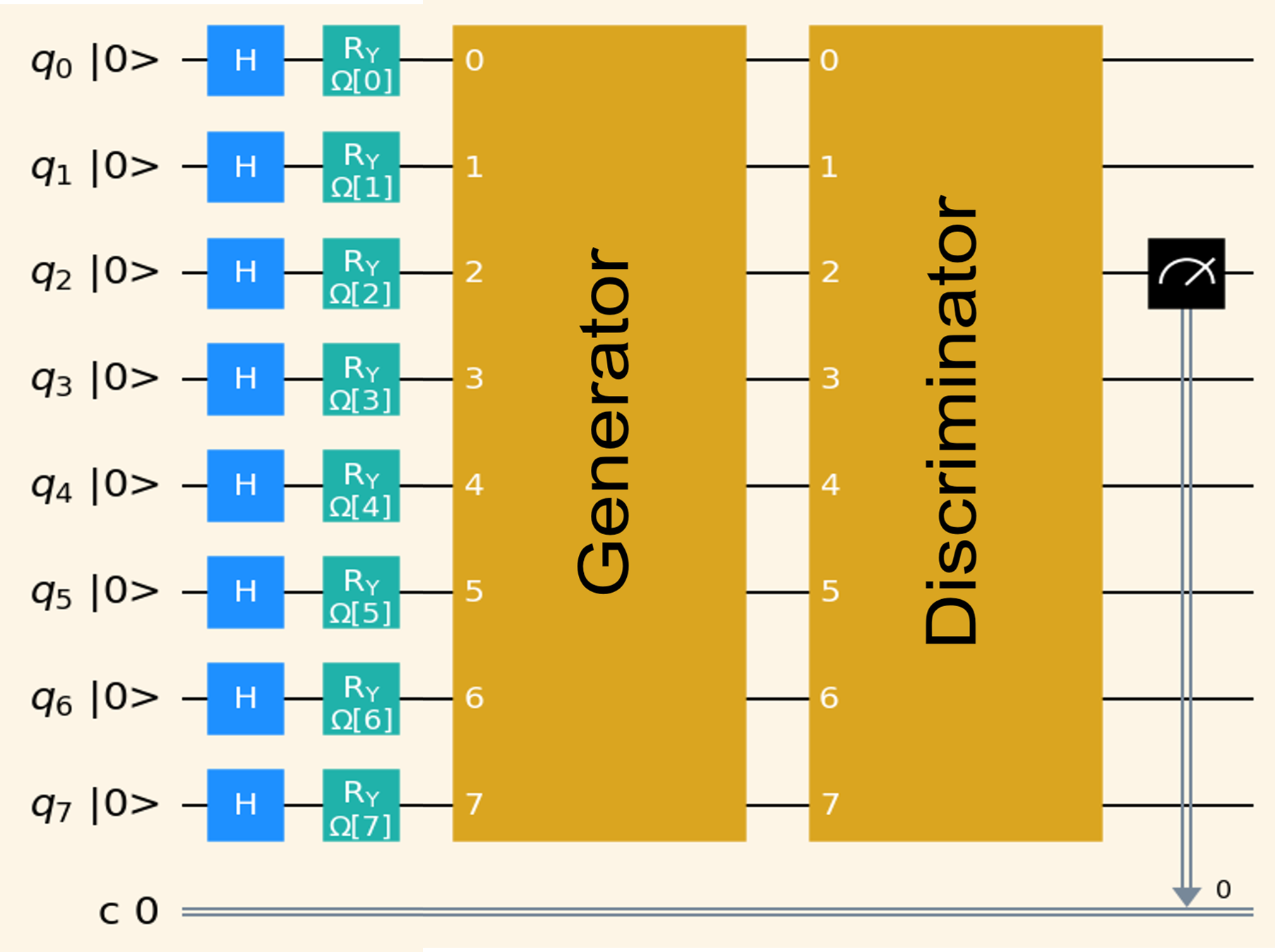} 
    \hspace{0.5cm}
    \includegraphics[height=0.32\textwidth]{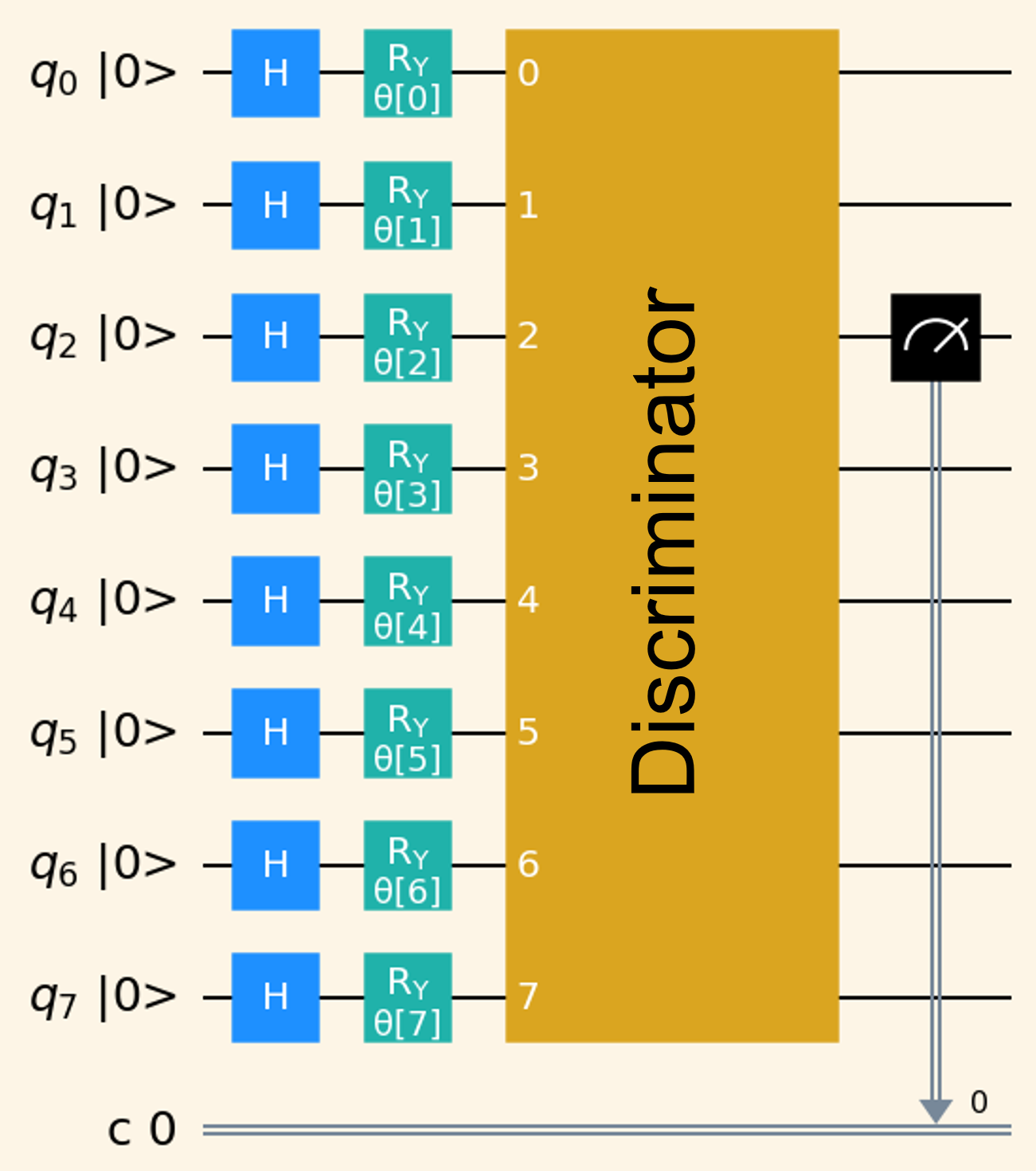}
    \caption{The full qGAN training circuits: (left) the circuit for generator and discriminator training with fake images. (right) The circuit for discriminator training with true images, where only the discriminator part and not the generator part is needed.}
    \label{fig:train_circuit}
\end{figure}

The full qGAN model \cite{QGAN} follows the adversarial training approach of classical GANs, described in Ref. \cite{goodfellow}. Since our model is fully quantum, it uses a quantum generator and a quantum discriminator circuit. The states of all qubits are initialized in the $\ket{0}$ basis state, followed by a Hadamard ($H$) gate to initiate superposition, as shown in figure \ref{fig:train_circuit} (left). Random Y-rotational ($RY$) gates with parameters $\Omega \in [-1,1]$ are used to generate noise and create individual images. The gate angles $\Omega$ are scaled by pixel standard deviations for accurate pixel energy variations, and a random factor between [-1/4, 1/4] accounts for primary particle energies.

\subsection{Training with Fake Images}
In the implementation of the full qGAN model, the two circuits -- generator and discriminator -- are directly connected to train on fake images or to train the generator. Both circuits use the same qubits and can be combined into one large circuit, as visualized in figure \ref{fig:train_circuit}. Therefore, no intermediate measurement is required between the generator and discriminator. This is an enormous advantage of the full qGAN model compared to hybrid approaches, since intermediate measurements can distort the result. This happens, for example, due to readout errors or due to the fact that the measurement operation is only an approximation to the probability vector of a quantum state instead of providing a precise result.

\begin{figure}[t!]  
    \center
    \includegraphics[width=0.45\textwidth]{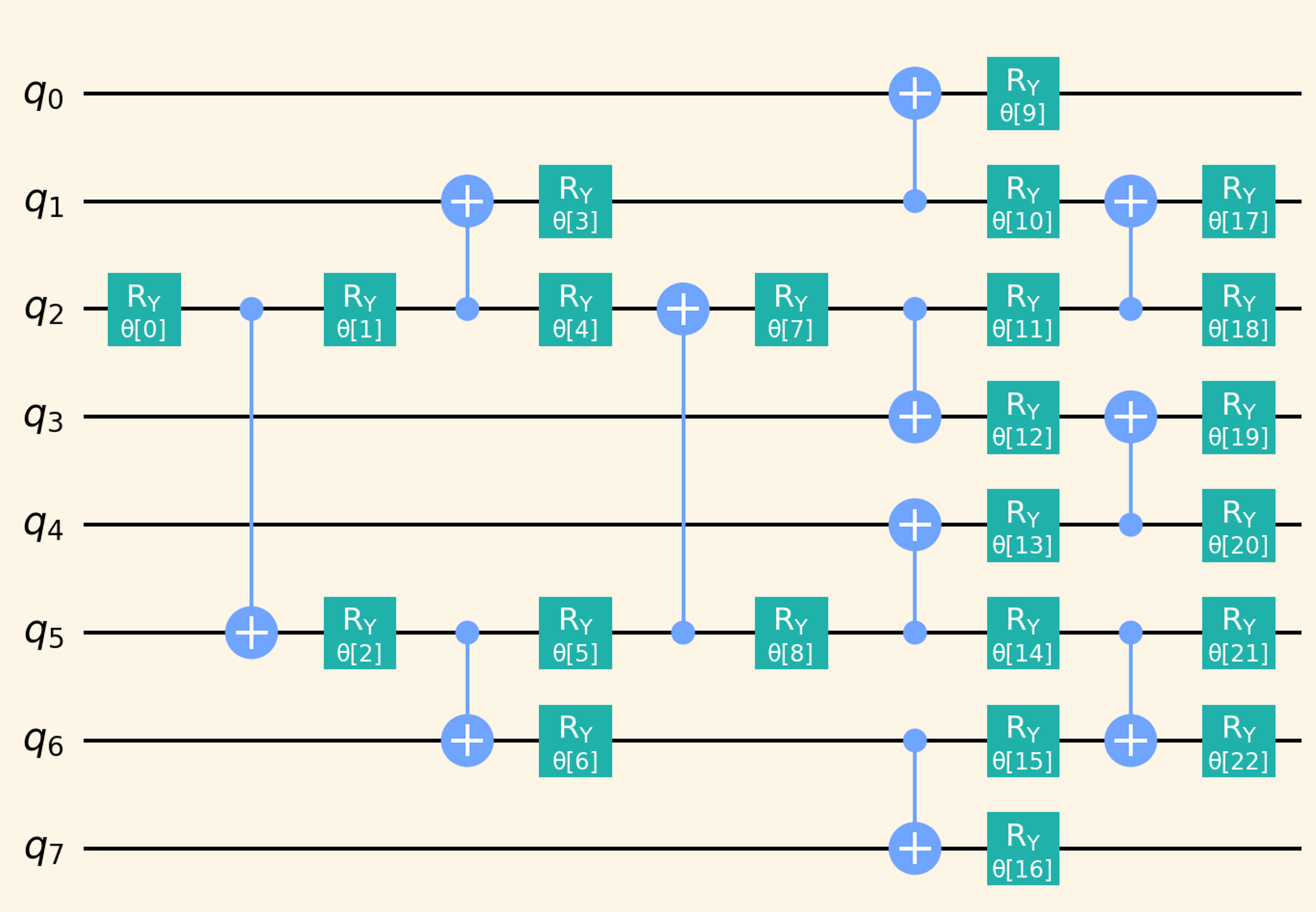} 
    \hspace{0.5cm}
    \includegraphics[width=0.45\textwidth]{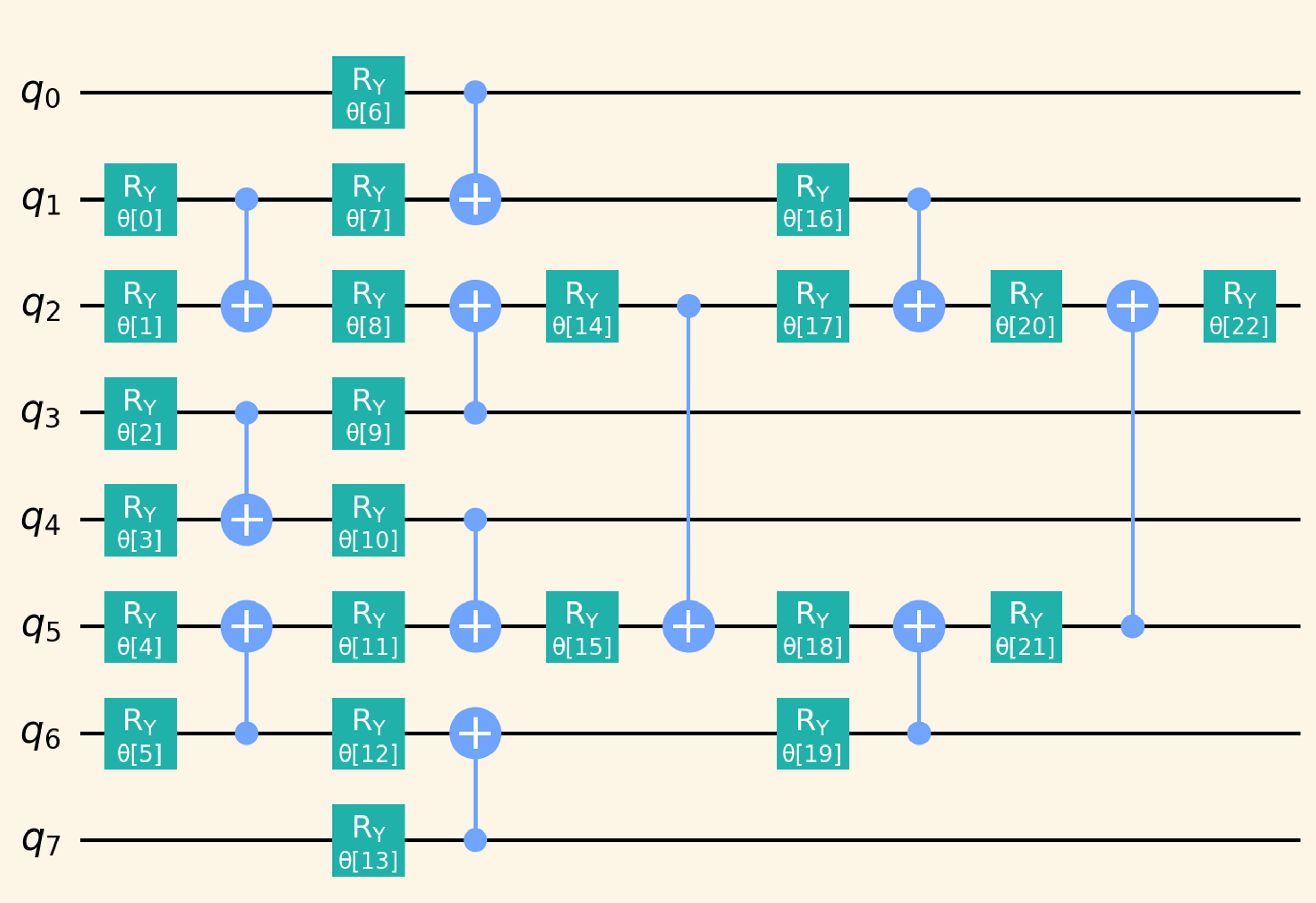}
    \caption{(left) The MERA-upsampling quantum circuit used as a generator and (right) the MERA-downsampling quantum circuit used as a discriminator.}
    \label{fig:circuits}
\end{figure}

The MERA-upsampling (MERA-up) circuit architecture is employed for the generator and the MERA-downsampling (MERA-down) architecture for the discriminator, as shown in figure \ref{fig:circuits}. Both circuit architectures are fractions from the Multi-scale Entanglement Renormalization Ansatz (MERA) introduced in Ref. \cite{TTN_MERA}. The main components for the circuits are two-qubit CX entangling gates to connect the qubits, and single qubit $RY$ gates with trainable parameters $\theta$.
In the MERA-up generator, the information from one qubit is upsampled to the others, and in the MERA-down discriminator, the information from multiple qubits is compressed or downsampled into one qubit, which is eventually measured and compared to classical data. This concept is similar to the one used in conventional machine learning generative models with neural networks, for example GANs \cite{goodfellow}.
The measurement result of the discriminator qubit corresponds to the True/Fake probability. When $\ket{0}$ is measured, the image corresponds to a fake image, and when $\ket{1}$ is measured, it corresponds to a true image. The True/Fake probability is obtained by measuring the discriminator output for multiple shots and averaging the result. The True/Fake loss is calculated by the binary cross entropy loss, as described in Ref. \cite{goodfellow, QGAN}. To confuse the discriminator, the labels of the true and fake images are swapped when training the generator.

\subsection{Training with True Images}
\begin{figure}[h!]  
    \center
    \includegraphics[height=0.3\textwidth]{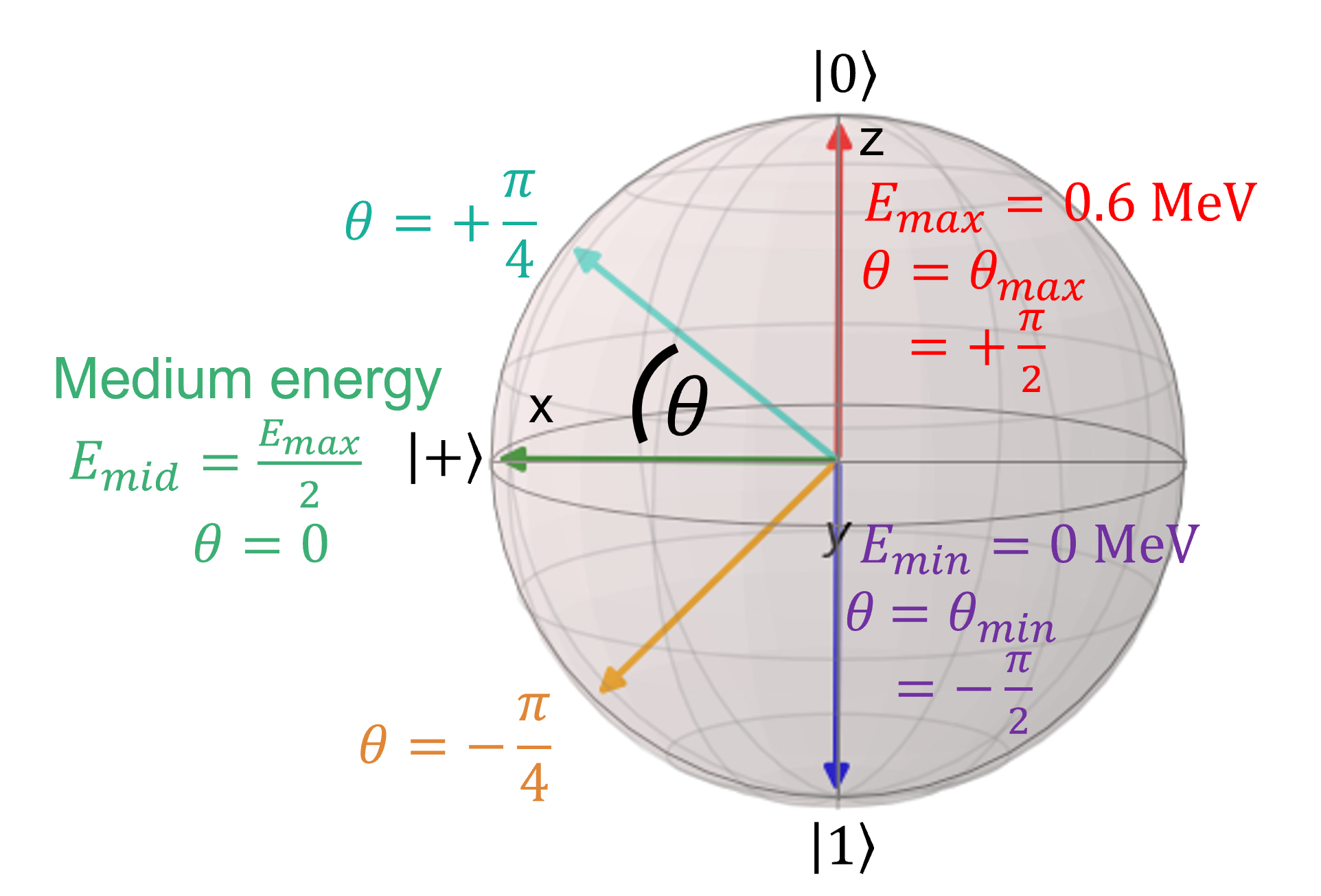} 
    \hspace{1.5cm}
    \includegraphics[height=0.3\textwidth]{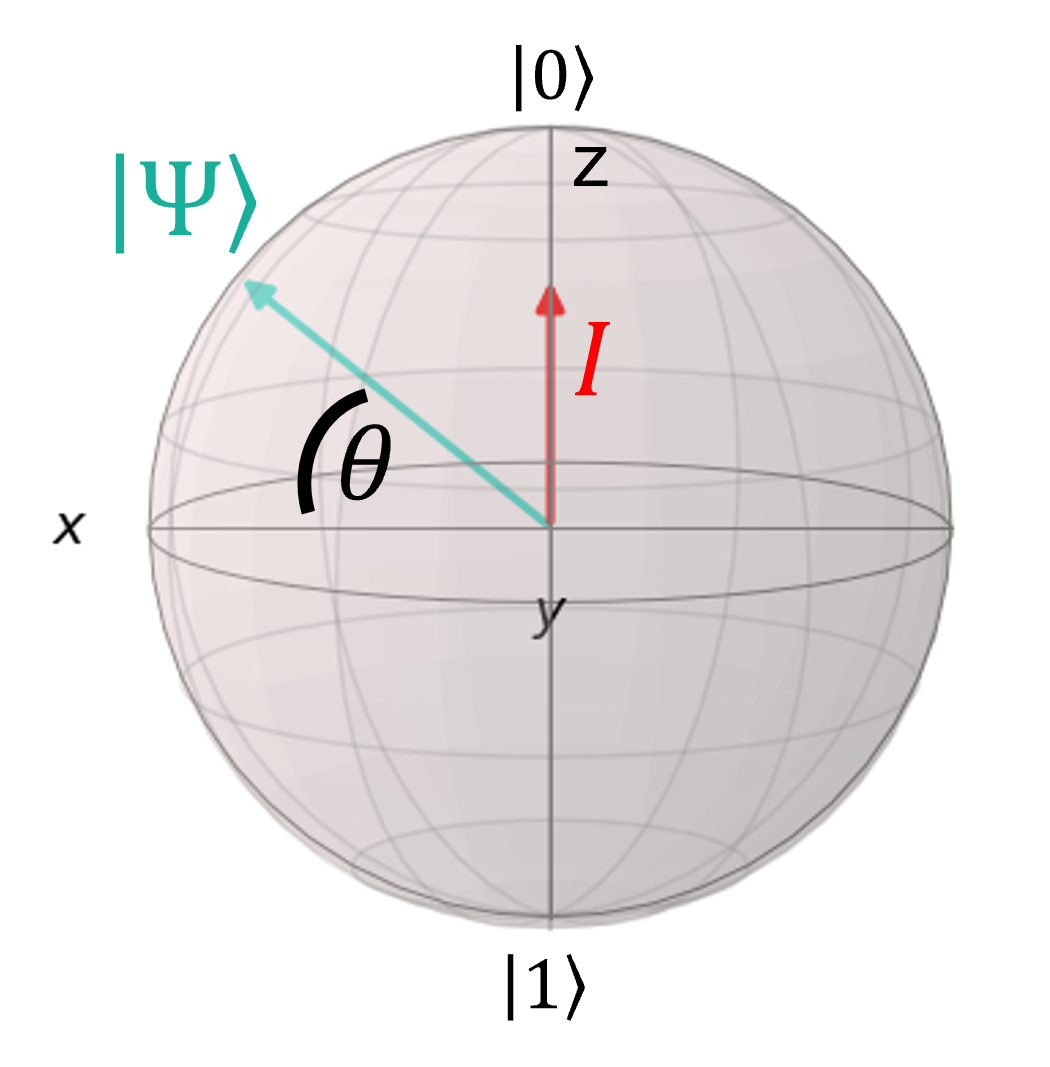}
    \caption{(left) The angle encoding from classical to quantum data and (right) the decoding.}
    \label{fig:encoding_decoding}
\end{figure}

To train the discriminator with true images, the generator is not required and the images from the classical training set must be encoded into quantum images. For this purpose, angle encoding with the rotation angles $\theta$ is employed, as visualized in the circuit in figure \ref{fig:train_circuit} (right). The angle encoding process for a single qubit is visualized in figure \ref{fig:encoding_decoding} (left) and described in the following. The minimum energy ($E_{min} = 0\,$MeV) is set to the $\ket{1}$ state at the bottom of the Bloch sphere and the maximum energy $E_{max} = 0.6\,$MeV is set to the $\ket{0}$ state at the top. The angle $\theta$ acts in the x-z-plane of the Bloch sphere and is defined as zero in the $\ket{+}$ state. 

For the encoding of the true images: first, a $H$-gate is applied to initiate superposition. Next, one $RY$-gate is placed. The rotational angle $\theta$ of the $RY$-gate correlates with the encoded pixel energy value. The transformation of the pixel energy $E$ into the angle $\theta$ is performed with:
\begin{equation}
    \theta = \frac{2 \cdot \theta_{max}}{E_{max}} \cdot E -  \theta_{max} \,.
    \label{eq:encoding}
\end{equation}

\subsection{Creating Shower Images for Inference}
For generating showers in the inference process, the outputs of all qubits of the generator circuit are measured directly to make them visible, and no discriminator is required. For decoding the measurements into energy values, the circuit must be executed for multiple shots and the number of $\ket{0}$ state measurements is counted. This allows the intersection $I$ of the vertical axis on the Bloch sphere (z-axis) to be calculated:
\begin{equation}
    I = \frac{\textrm{counts}(\ket{0})}{\textrm{number of shots}}.
\end{equation}
The calculated intersection $I$ is transformed to the angle $\theta$ using trigonometric functions by: $\theta = \textrm{arcsin(}I\textrm{)}$. The figure \ref{fig:encoding_decoding} (right) shows an example state $\Psi$ with its intersection $I$ and the corresponding angle $\theta$. The back transformation of the angle $\theta$ into an energy is done with resolving equation \ref{eq:encoding} for the energy $E$:
\begin{equation}
    E = \frac{\theta-\theta_{max}}{2 \cdot \theta_{max}} \cdot E_{max} .
\end{equation}

\section{The Adversarial Training of the Full Quantum GAN Model}
Separate generator and discriminator learning rates are employed by empirical experience: the generator learning rate 0.02 and the discriminator learning rate 0.04 are chosen. The training converges much faster when the discriminator is trained more frequently than the generator. A similar observation was made for a hybrid qGAN model in Ref. \cite{1QGAN_hep}. In the full qGAN model, the discriminator is trained five times at each optimization step, followed by one generator training. The model is trained for 1\,000 epochs, where one epoch contains eight optimization steps (batch size of eight). The generator and discriminator employ two distinct exponential learning rate decays: the learning rate decay for the generator is 0.006 and for the discriminator 0.007. Training and inference use 1\,024 shots. The hyperparameters are found by hyperparameter searches using the Optuna \cite{optuna} library and empirical testing. All experiments in this paper are conducted using quantum simulators.

Two-sided label smoothing is applied during training: true labels 1 are smoothed to 0.9 and fake labels 0 to 0.1. This improves training convergence and the achieved accuracy. All rotational parameters are initialized with zeros to provide a stable training start point. The SPSA optimizer \cite{SPSA} is used for calculating the parameter update.

\subsection{Training Evaluation of the Full Quantum GAN}
\begin{figure}[ht!]  
    \center
    \includegraphics[height=0.35\textwidth]{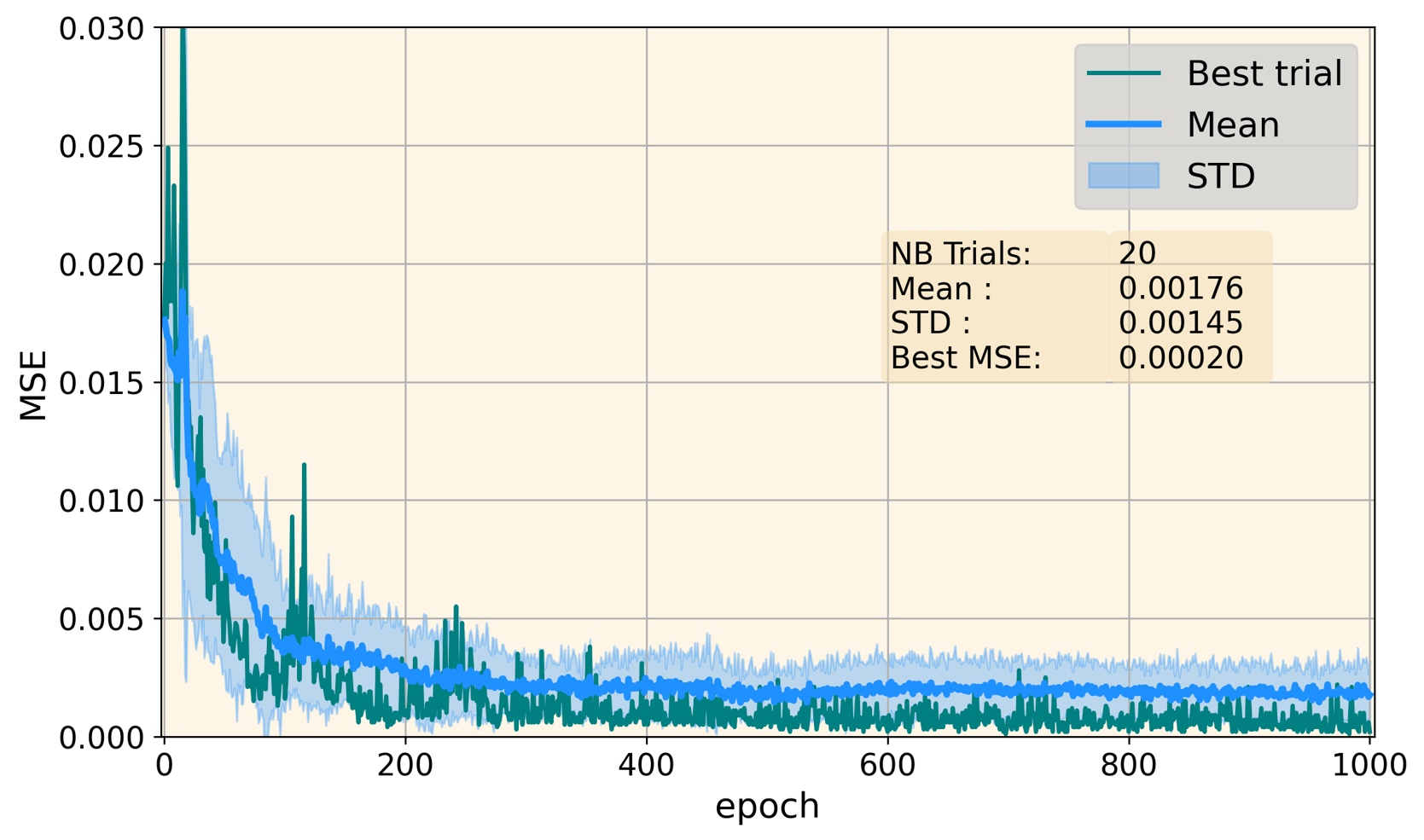} 
    \hspace{0.5cm}
    \includegraphics[height=0.4\textwidth]{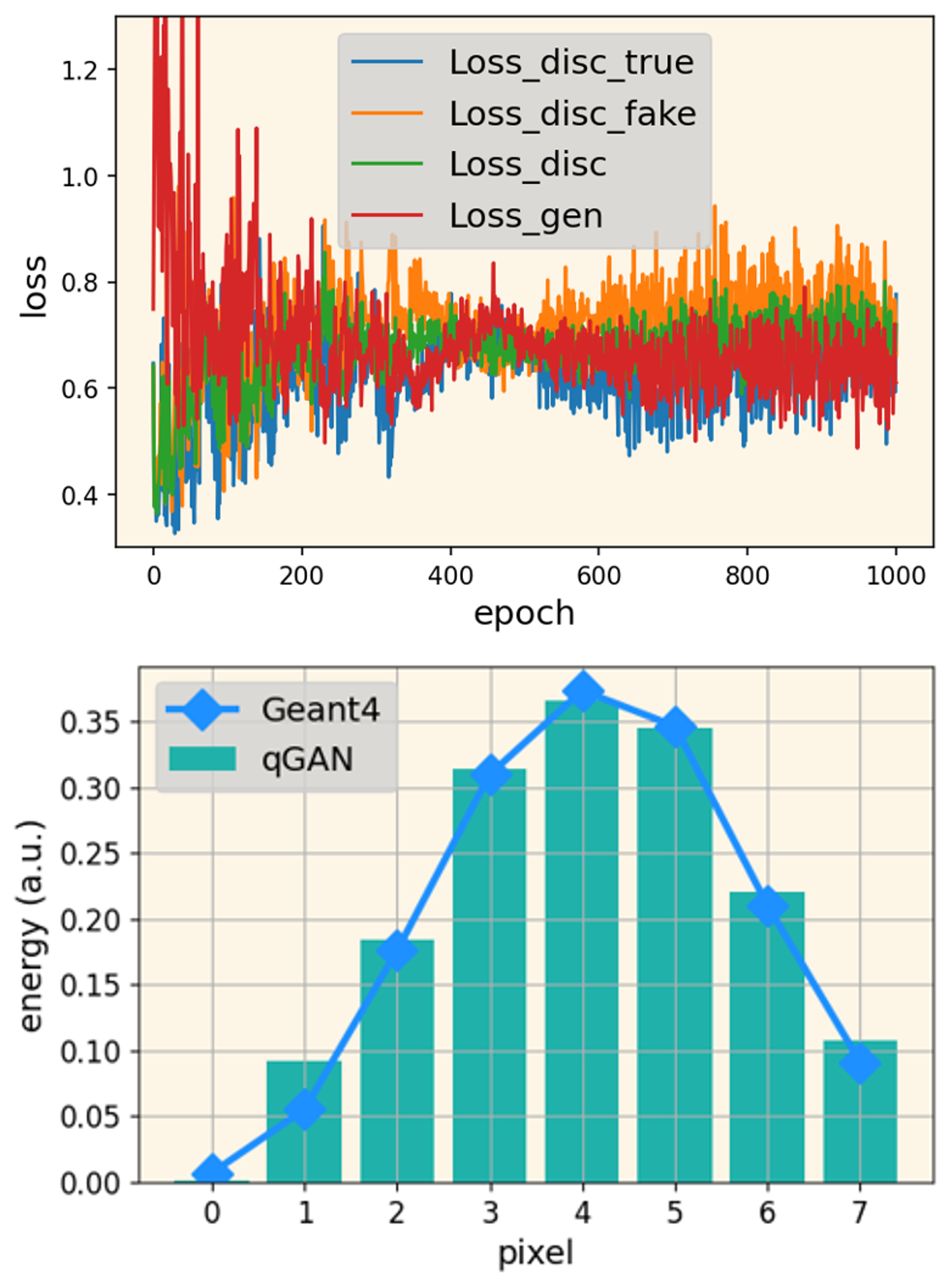} 
    \caption{(left) The MSE accuracy metric is plotted as a function of the training epochs for 20 trials. (top right) The characteristic training losses of the best trial. (bottom right) The generated average shower image in inference. The box shows the mean MSE, its standard deviation (STD) and the MSE of the best trial of the last training epoch.}
    \label{fig:qgan_training}
\end{figure}

Figure \ref{fig:qgan_training} (left) visualizes the statistics of 20 training trials (repeating the training with the same hyperparameters) with the average MSE as the accuracy metric. The MSE is computed pixel by pixel between 50 generated and training images. Due to the hyperparameter optimization, the mean value (blue solid line) converges quickly and remains at a stable level. The standard deviation (light blue band) is narrow, and the best trial (green solid line) shows little oscillations. All these factors indicate a smooth, stable and successful training. 

Next, the characteristic training losses of the best trial are shown in figure \ref{fig:qgan_training} (top right). Plotted are: the generator loss (red), the discriminator true loss (blue), the discriminator fake loss (orange) and the summed discriminator losses (green). Due to the zero initialization of the circuit parameters, the composite discriminator loss is lower than the generator loss in the first epochs. In the following epochs, the losses converge to increasingly similar values, resulting in a good training convergence. Moreover, the losses remain stable towards the end of the training due to the learning rate decay. Between the single epochs, the losses oscillate in a larger range, likely due to the small number of batches. A larger number of batches per epoch would potentially smooth the losses.

\subsection{Inference Evaluation Quantum GAN}
To evaluate the inference accuracy, the same number of images is generated as in the Geant4 test set. As a first accuracy metric, the average shower image of the best trial is provided in figure \ref{fig:qgan_training} (bottom right). It can be observed that the full qGAN model can accurately reproduce the average shower image of Geant4 with an overall MSE of: MSE $= 1.77 \pm 1.46$.

\section{Hybrid Quantum-Classical Model for Comparison}
\label{sec:hybrid_angle_qGAN}
In a final test, a hybrid qGAN combining the MERA-up quantum generator with a classical discriminator is investigated. This comparison could enable to observe possible advantages of the quantum behavior over the classical behavior, or vice versa. The classical discriminator is constructed with fully connected layers. Three discriminator sizes S, M and L are tested: small (S: $153$ trainable parameters), medium (M: $433$ trainable parameters) and large (L: $1\,889$ trainable parameters). 
The hyperparameters of all the hybrid qGAN model that achieve training convergence are: the generator learning rate 0.01, the discriminator learning rate 0.006 and a joint learning rate decay 0.006. The training comprises 1\,000 epochs, each containing one optimization step. The generator and discriminator are equally optimized once per training step.

\subsection{Training Evaluation Hybrid qGAN}
\begin{figure}[t!]  
    \center
    \includegraphics[width=0.6\textwidth]{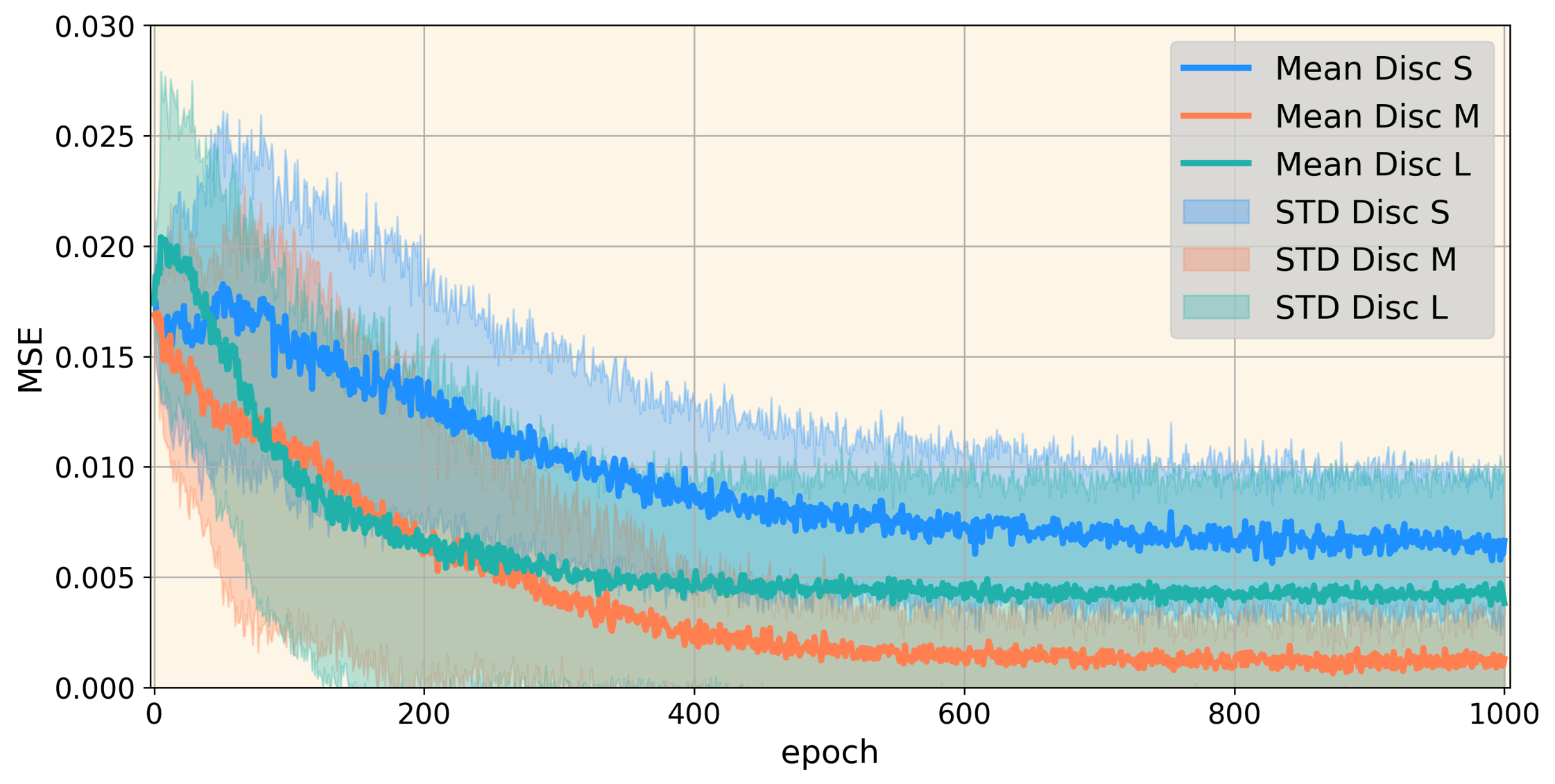} 
    \includegraphics[width=.38\textwidth]{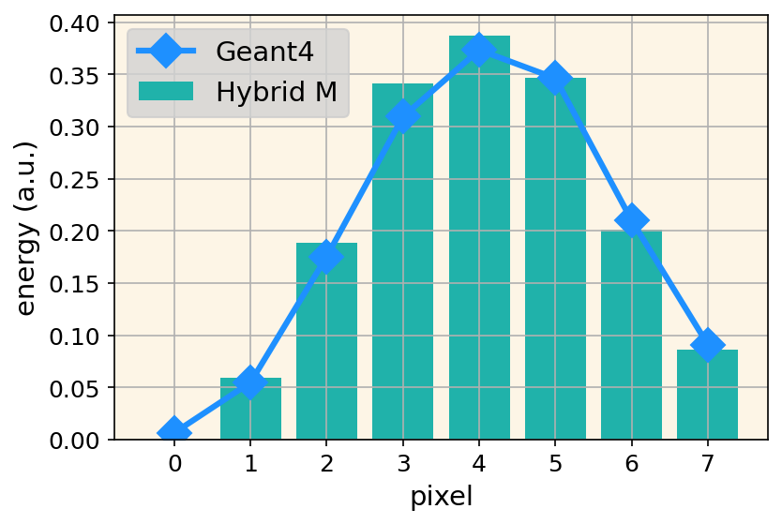}
     \caption{(left) The MSE is plotted as an accuracy metric as a function of the training epochs for the three hybrid qGAN models. (right) The average shower shape plot generated by the hybrid M qGAN model of the best trial.}
    \label{fig:hybrid_qgan_training_statistics}
\end{figure}

Ten training trials are performed for each of the three models, and shown in figure \ref{fig:hybrid_qgan_training_statistics} (left). The average MSE converges for all models to stable results: towards the end of the training, the hybrid M model converges to the lowest MSE (MSE $= 1.50 \pm 2.15$), which corresponds to the best accuracy. The reason why the S (MSE $= 6.65 \pm 3.11$) and L (MSE $= 3.80 \pm 4.32$) models perform worse is due to the GAN principle: in the adversarial training, the best accuracy is obtained when the generator and discriminator model are balanced. Because the L (S) discriminator outperforms (underperforms) the generator, the accuracy becomes worse. The M model exhibits a representational power that matches best the one of the generator. 

\subsection{Inference Evaluation Hybrid qGAN}
Figure \ref{fig:hybrid_qgan_training_statistics} (top right) displays the average shower image for the best hybrid M model. It agrees well with the Geant4 prediction. 
Since the accuracy of the full qGAN model (MSE $= 1.77 \pm 1.46$) is very similar to the best hybrid model M (MSE $= 1.50 \pm 2.15$), the representational power of the two discriminator models, the full quantum and the hybrid, can be considered approximately similar within the given statistics. This means that the MERA-down discriminator quantum circuit architecture with only 20 trainable parameters performs equally well to a classical neural network with more than 400 trainable parameters. This rises the question if variational quantum circuits requiring far fewer trainable parameters are able to achieve the same representational power as neural networks? 
Similar results are observed in related studies. For example, in Ref. \cite{QNN_capacity} and Ref. \cite{QDNN_power}, it is claimed that quantum circuits have a higher storage and modelling capacity as well as a better computational efficiency than neural networks because they employ quantum principles.
It should be noted that both the full qGAN and the hybrid qGAN models are unable to reproduce the more sophisticated physics accuracy metrics beside the average shower image, for example the pixel-wise correlation patterns.

\section{Discussion and Conclusion}
\label{sec:fullqgan_conclusion}
In this paper, a full qGAN model was developed and trained as a proof of concept. The trained model was able to accurately reproduce the average shower image. 

For comparison a hybrid qGAN was trained with three discriminator sizes in terms of number of parameters. Only the hybrid model with the medium discriminator size was able to achieve similar MSE values as the full qGAN. Training with the small and large discriminator produced worse results due to an imbalance of the power between the generator and discriminator models. 

It is worth noting, that the full qGAN model with only 20 parameters accomplished a statistically similar good accuracy than the hybrid model M with 433 parameters. This result encourages to pursue this approach further towards larger models and increasing statistics. Future research should also conduct a more thorough examination of the model's accuracy, including aspects such as pixel-wise correlations, to gain deeper insights.

\section*{Acknowledgment}
This work has been sponsored by the Wolfgang Gentner Programme of the German Federal Ministry of Education and Research and by the CERN Quantum Technology Initiative.

\section*{References}
\bibliographystyle{IEEEtran}
\bibliography{main}

\begin{thebibliography}{10}
\providecommand{\url}[1]{#1}
\csname url@samestyle\endcsname
\providecommand{\newblock}{\relax}
\providecommand{\bibinfo}[2]{#2}
\providecommand{\BIBentrySTDinterwordspacing}{\spaceskip=0pt\relax}
\providecommand{\BIBentryALTinterwordstretchfactor}{4}
\providecommand{\BIBentryALTinterwordspacing}{\spaceskip=\fontdimen2\font plus
\BIBentryALTinterwordstretchfactor\fontdimen3\font minus
  \fontdimen4\font\relax}
\providecommand{\BIBforeignlanguage}[2]{{%
\expandafter\ifx\csname l@#1\endcsname\relax
\typeout{** WARNING: IEEEtran.bst: No hyphenation pattern has been}%
\typeout{** loaded for the language `#1'. Using the pattern for}%
\typeout{** the default language instead.}%
\else
\language=\csname l@#1\endcsname
\fi
#2}}
\providecommand{\BIBdecl}{\relax}
\BIBdecl

\bibitem{variational_quantum_circuits}
\BIBentryALTinterwordspacing
M.~Vogel, ``{Quantum Computation and Quantum Information, by M.A. Nielsen and
  I.L. Chuang},'' \emph{Contemporary Physics - CONTEMP PHYS}, vol.~52, pp.
  604--605, 11 2011. [Online]. Available:
  \url{https://profmcruz.files.wordpress.com/2017/08/quantum-computation-and-quantum-information-nielsen-chuang.pdf}
\BIBentrySTDinterwordspacing

\bibitem{qGAN_example}
\BIBentryALTinterwordspacing
L.~Hu, S.-H. Wu \emph{et~al.}, ``Quantum generative adversarial learning in a
  superconducting quantum circuit,'' \emph{Science advances}, vol.~5, 2019.
  [Online]. Available: \url{https://www.science.org/doi/10.1126/sciadv.aav2761}
\BIBentrySTDinterwordspacing

\bibitem{qGAN_example2}
\BIBentryALTinterwordspacing
P.-L. Dallaire-Demers \emph{et~al.}, ``Quantum generative adversarial
  networks,'' \emph{Physical Review A}, vol.~98, 2018. [Online]. Available:
  \url{https://doi.org/10.1103\%2Fphysreva.98.012324}
\BIBentrySTDinterwordspacing

\bibitem{calorimeter_cern}
C.~Fabjan and F.~Gianotti, ``{Calorimetry for Particle Physics},''
  \emph{Reviews of Modern Physics}, 2003.

\bibitem{quantum_data_set}
\BIBentryALTinterwordspacing
F.~Rehm, ``{Downsampled Calorimeter Shower Images to 8 Pixels},'' 2021.
  [Online]. Available: \url{https://doi.org/10.5281/zenodo.7025233}
\BIBentrySTDinterwordspacing

\bibitem{CLIC_dataset}
\BIBentryALTinterwordspacing
M.~Pierini and M.~Zhang, ``{CLIC Calorimeter 3D images: Electron showers at
  Fixed Angle},'' Jan. 2020. [Online]. Available:
  \url{https://doi.org/10.5281/zenodo.3603122}
\BIBentrySTDinterwordspacing

\bibitem{Geant4}
\BIBentryALTinterwordspacing
S.~Agostinelli \emph{et~al.}, ``{{GEANT4--a simulation toolkit}},'' vol. 506,
  2003, pp. 250--303. [Online]. Available:
  \url{https://inspirehep.net/files/6c9c0b62bbc8dc0401fca11a5fe5c87c}
\BIBentrySTDinterwordspacing

\bibitem{state_preparation2}
\BIBentryALTinterwordspacing
M.~Weigold, J.~Barzen \emph{et~al.}, ``{Data Encoding Patterns for Quantum
  Computing},'' in \emph{Proceedings of the 27th Conference on Pattern
  Languages of Programs}, ser. PLoP '20.\hskip 1em plus 0.5em minus 0.4em\relax
  USA: The Hillside Group, 2020. [Online]. Available:
  \url{https://hillside.net/plop/2020/papers/weigold.pdf}
\BIBentrySTDinterwordspacing

\bibitem{QGAN}
\BIBentryALTinterwordspacing
S.~Lloyd and C.~Weedbrook, ``Quantum generative adversarial learning,''
  \emph{Phys. Rev. Lett.}, vol. 121, p. 040502, Jul 2018. [Online]. Available:
  \url{https://link.aps.org/doi/10.1103/PhysRevLett.121.040502}
\BIBentrySTDinterwordspacing

\bibitem{goodfellow}
\BIBentryALTinterwordspacing
I.~J. Goodfellow, J.~Pouget-Abadie \emph{et~al.}, ``{Generative Adversarial
  Networks},'' 2014. [Online]. Available: \url{https://arxiv.org/abs/1406.2661}
\BIBentrySTDinterwordspacing

\bibitem{TTN_MERA}
\BIBentryALTinterwordspacing
E.~Grant \emph{et~al.}, ``Hierarchical quantum classifiers,'' \emph{npj Quantum
  Information}, vol.~4, no.~1, dec 2018. [Online]. Available:
  \url{https://doi.org/10.1038\%2Fs41534-018-0116-9}
\BIBentrySTDinterwordspacing

\bibitem{1QGAN_hep}
\BIBentryALTinterwordspacing
F.~Rehm \emph{et~al.}, ``{Quantum Machine Learning for {HEP} Detector
  Simulations},'' 2021. [Online]. Available:
  \url{http://ceur-ws.org/Vol-3041/363-368-paper-67.pdf}
\BIBentrySTDinterwordspacing

\bibitem{optuna}
\BIBentryALTinterwordspacing
T.~Akiba \emph{et~al.}, ``{Optuna: A Next-generation Hyperparameter
  Optimization Framework},'' 2019. [Online]. Available:
  \url{https://optuna.org/}
\BIBentrySTDinterwordspacing

\bibitem{SPSA}
\BIBentryALTinterwordspacing
A.~H. Alhabsi, ``Improved {SPSA} optimization algorithm requiring a single
  measurement per iteration,'' in \emph{10th International Conference on
  Information Science, Signal Processing and their Applications}, 2010.
  [Online]. Available: \url{https://ieeexplore.ieee.org/document/5605476/}
\BIBentrySTDinterwordspacing

\bibitem{QNN_capacity}
\BIBentryALTinterwordspacing
R.~Zhao and S.~Wang, ``{A review of Quantum Neural Networks: Methods, Models,
  Dilemma},'' 2021. [Online]. Available: \url{https://arxiv.org/abs/2109.01840}
\BIBentrySTDinterwordspacing

\bibitem{QDNN_power}
\BIBentryALTinterwordspacing
C.~Zhao and X.-S. Gao, ``{QDNN: DNN with Quantum Neural Network Layers},''
  2019. [Online]. Available: \url{https://arxiv.org/abs/1912.12660}
\BIBentrySTDinterwordspacing

\end{thebibliography}

\end{document}